# Informed recruitment or the importance of taking stock

Meritxell Genovart[1,3], Roger Pradel[2], Remi Choquet[2] and Daniel Oro[1,3]

[1] CEAB (CSIC), Acces Cala Sant Francesc 14, 17300 Blanes, Catalonia, Spain

[2] CEFE, CNRS, Univ. Montpellier, Univ. Paul Valéry Montpellier 3, EPHE, IRD, Montpellier, France

[3] IMEDEA (CSIC-UIB), Miquel Marques 21, 07190 Esporles, Spain

Author for correspondence:

Meritxell Genovart

IMEDEA (CSIC-UIB)

Miquel Marques 21

07190 Esporles

Spain

e-mail: m.genovart@uib.es

One important decision, especially for long-lived species, is where and when to start reproduction, i.e. where and when to recruit. For an informed decision on recruitment, individuals may prospect potential breeding sites hence reducing uncertainty. Recruitment may also be delayed or advanced over a range of sexually mature ages and this individual decision may influence survival costs of first reproduction, senescence patterns and ultimately lifetime reproductive success. Here we show that recruitment in the long-lived Audouin's gull occurs only when the individual was present on the colony before. This *sine qua non* condition for recruitment suggests that taking stock of the breeding site is essential for making informed decisions with important consequences for fitness prospects.

## 1. Introduction

Prospecting is a behavioural process of spatial exploration addressed for gathering information about the features of heterogeneous environments and reducing their inherent unpredictability. This process (that has also been termed scouting) has been described for many species, from ants and bees to seabirds and pinnipeds. From an evolutionary point of view, prospecting is important because it influences the decision made by an individual about where to breed and thus its fitness prospects (e.g. survival costs of breeding, strength of senescence and ultimately lifetime reproductive success). For instance, once an individual attains its sexual maturity it has to take stock where to start reproduction, i.e. where to recruit. Most often, dispersal between patches (either natal or breeding dispersal) is a risky decision compared to a philopatric strategy because dispersing individuals face the challenges of the unknown [1,2].

To reduce the risks of dispersal, individuals prospect to gather information about the quality of the different patches and make what has been termed *informed dispersal* [3,4]. Prospecting is particularly crucial when perturbations affect a patch and informed dispersal to other patches is a resilient mechanism to overcome environmental stress [1,5]. Apart of where to breed, prospecting provides information on other important decisions, such as when to start breeding. This occurs particularly for long-lived species, because recruitment varies with age and can be deferred to old ages. Recruitment decision about when to incorporate to the breeding compartment of the population can have similar evolutionary importance than the decision made about where to breed, because age of first breeding influences survival costs of reproduction and lifetime reproductive success [6–8].

In this letter, we use a long-term individual monitoring dataset on a long-lived social seabird to test the hypothesis that individuals prospect the natal breeding site the years before recruiting and whether this pattern changes with recruitment age.

## 2. Material and methods

**(a) Data acquisition**

Audouin's gull is a long-lived colonial seabird specialized in nocturnal feeding on small pelagics but also able to exploit anthropogenic foraging resources. Its breeding season (from settlement to incubation and chick fledgling) lasts approximately 3 months. Modal clutch size is three eggs. After breeding, birds migrate to wintering sites. The study was performed at the colony of Punta de la Banya, at the Ebro Delta in the western Mediterranean. The site has held the world's largest colony of the species for more than two decades, with a maximum of 73% of the global population. During 1995-2010, 17877 chicks were individually marked with plastic rings, which can be resighted from the distance using spotting scopes. The breeding status was ascertained for gulls whose breeding behaviour was monitored. The criteria for assessing that a gull was breeding ranged from uncertain behaviours (e.g. bringing branches for nest building, copulation, courtship feeding) to certain behaviours (e.g. incubating, feeding chicks). From those chicks marked, 7907 different individuals were resighted at least once at the study site.

**(b) Statistical analyses**

Our capture-mark-recapture (CMR) model considers that there are potentially two milestones in the accession to reproduction at the Ebro delta colony for Audouin's gulls born locally: the first consists of taking stock of the colony by visiting it during one reproductive season after fledging; the second milestone is crossed when the individual actually breeds for the first time in the colony. Individuals can thus be in 3 different sequential states as represented in figure 1: not recruited uninformed ($NR_U$)), informed but not recruited ($NR_I$), recruited (R). The main aim of the study is to assess whether the informed intermediate state is required or whether accession to reproduction can be effected directly from the uninformed state. The individuals may also die at some point, thus joining the absorbing dead state. Apart from the transitions describing the process of recruitment (electronic supplementary material, S1), the model additionally includes survival parameters, and, detection not being perfect, detection parameters.

We tested that hypothesis within the multievent CMR framework [9]. In accordance with previous findings [10], survival was modelled as age-dependent up to age 6 and detection was modelled as time-dependent. Also, when running the goodness-of-fit tests on the histories of presence of individuals at least 3 year old, there is strong evidence of trap dependence among breeders (Test 2.CT: $X^2_{10}$=197.16, P < 10$^{-12}$), which means that detection differs markedly between individuals seen and not seen the year before. Although we have no clear explanation for this phenomenon (some individuals may breed less regularly or may occupy less visited part of the colony), we clearly need to take it into account. Thus we estimated separate detection parameters for the two kinds of individuals. We have also evidence of survival heterogeneity (Test 3.SR:

$X^2_{11}$=65.91, P < 10$^{-6}$). This is treated by allowing for different survival of 3, 4 and 5+ year-olds in the model. The remaining lack-of-fit is treated by means of a variance inflation factor of 1.35 (see appendix for details), which serves to scale up precision estimates and which enters the calculation of the quasi Akaike information criterion (QAICc) used to rank the models (see below). Beyond detection, the observation process includes an attempt at assessing whether the spotted individual is actually breeding at the time. Different clues leading to distinguish more or less strong evidence of breeding (see above), the model eventually includes parameters associated to the likelihood of each of these 5 events.

Model selection was limited to the points that had not been examined in previous studies and thus concerns the recruitment process. We notably checked whether individuals breeding for the first time should be assimilated to breeders seen or not seen the previous year regarding detection. If the lower detection of individuals not seen the previous year corresponds to a probability of absence, then the probability of detection of breeders seen the previous year is clearly more appropriate as individuals breeding for the first time on the colony are obviously present on the colony. If it corresponds to presence on a less accessible part of the colony, we do not know in advance which detection probability would be more appropriate. Regarding the process of recruitment itself, we examined different age structures. Sexual maturity is reached only at age 3. We first examined whether the probability of recruitment increased from age 3 to age 4 and then remained constant. Then, we considered the possibility that it increased again from age 4 to age 5 and stabilized. Eventually, we considered the possibility that recruitment increased again from age 5 to age 6 before stabilizing at that age. Models were compared by means of the quasi Akaike information criterion that, unlike the original Akaike information criterion, is corrected for the residual lack of fit by incorporating the variance inflation factor in its calculation. All models were fitted with program E-SURGE version 2.1.4 [11].

**3. Results and discussion**

We first show that recruitment in Audouin's gull at the study site occurs at a very fast rate: individuals start breeding at 3y old and recruitment ends for individuals aged 6y, the two first age-classes (3y and 4y old) showing much higher recruitment probabilities (electronic supplementary material, S2 and figure 1). Second and importantly, our results clearly show that most Audouin's gulls do not recruit to their natal site without prospecting in advance and that this pattern does not vary with age (figure 1). While recruitment for most age classes never occurs without having prospecting the colony during the precedent years, individuals recruiting at the oldest age-class recruit without prospecting, but at very low probabilities. Results highlight the importance of prospecting to gather information for reducing uncertainty about an important decision (where and when to breed) with fitness consequences. Even though we do not specifically test for prospecting more than one year before recruitment, the rapid recruitment with age recorded here for the study species suggests that gulls do not mostly need to prospect for more than one breeding season. For other long-lived seabirds with slower life histories, such as Procellaridae and Sulidae , prospecting can last longer over a period of several years before a decision is made about where and when to recruit [12,13]. Empirical studies also show that individuals may prospect several breeding patches to gather information at relatively large spatial scales [14,15]. Even though we cannot analyse resights of birds at other breeding sites due to the small sample size, several gulls were also observed at these sites before recruiting at the Ebro Delta, and this frequency declined with distance [16,17]. Some gulls were observed at several different sites, including the Ebro Delta and sites more than 300km away, within the same breeding season. If prospection occurs at different patches (occupied or empty) over time to make the decision about where to recruit in a later year, then some kind of temporal autocorrelation in patch suitability must occur for prospecting being selected [18–20]. The same should occur for breeding dispersal when individual has to decide between stay in the patch where it breeds and leaving this patch to reproduce elsewhere. Information gathering is also important for short-lived species, even though the prospecting phase for breeding habitat selection should last shorter both before recruitment and for later breeding dispersal events [21,22].

Our results confirm empirical evidences and theoretical predictions made about the importance of gathering information before taking stock for important decisions with fitness consequences. At individual level, some field studies show that individuals showing a more active and more suitable prospection behaviour show higher breeding success in subsequent years [23,24]. Theoretical models of decision making theory show that fitness prospects increase for individuals making suitable prospecting compared to individual having less prospecting personalities [25]. In social species, social information is used to make these decisions at a faster rate, because some information is readily available (e.g. the number of conspecifics and heterospecifics), and this may have consequences for population and community dynamics [26,27]. For instance, some individual-based models show that information use reduces patch extinction probabilities as long as information gathering increases [28]. Finally, prospecting and taking stock of recruitment and dispersal is also important for colonization of new patches and thus for the dynamics of metapopulations and spatially structured populations. Here, social information is not available and thus private information becomes crucial for making risky decisions when occupying an empty site [1]. More empirical and theoretical studies are needed to explore how prospecting processes occur when social information is not available, especially when the occupied patches are perturbed (e.g. predation, habitat loss) and population persistence depends on colonization rates.

Ethics. Our research did not require an ethical assessment prior to being conducted.

Data accessibility. Data will be available via CSIC repository under acceptation (https://digital.csic.es/).

Authors' contributions. M.G., R.P. and D.O. conceived the study; M.G. analysed the data with contributions from R.P.; M.G., R.P. and D.O. wrote the manuscript. All authors approved the final version of the manuscript and agree to be held accountable for the content therein.

Competing interests. We declare we have no competing interests.

Funding. Funding came from the Spanish Ministry of Science and project PICS (CSIC-CNRS).

Acknowledgements. We are grateful to all people helping with fieldwork over the years.

**References**

1. Payo-Payo A, Genovart M, Sanz-Aguilar A, Greño JL, García-Tarrasón M, Bertolero A, Piccardo J, Oro D. 2017 Colonisation in social species: the importance of breeding experience for dispersal in overcoming information barriers. *Sci. Rep.* **7**, 42866. (doi:10.1038/srep42866)

2. Forbes LS, Kaiser GW. 1994 Habitat choice in breeding seabirds: when to cross the information barrier. *Oikos* **70**, 377–384. (doi:10.2307/3545775)

3. Ponchon A, Garnier R, Grémillet D, Boulinier T. 2015 Predicting population responses to environmental change: the importance of considering informed dispersal strategies in spatially structured population models. *Divers. Distrib.* **21**, 88–100. (doi:10.1111/ddi.12273)

4. Reed JM, Boulinier T, Danchin E, Oring LW. 1999 Informed Dispersal. In *Current Ornithology*, pp. 189–259. Springer, Boston, MA. (doi:10.1007/978-1-4757-4901-4_5)

5. Ponchon A, Iliszko L, Grémillet D, Tveraa T, Boulinier T. 2017 Intense prospecting movements of failed breeders nesting in an unsuccessful breeding subcolony. *Anim. Behav.* **124**, 183–191. (doi:10.1016/j.anbehav.2016.12.017)

6. Fay R, Barbraud C, Delord K, Weimerskirch H. 2016 Variation in the age of first reproduction: different strategies or individual quality? *Ecology* **97**, 1842–1851. (doi:10.1890/15-1485.1)

7. Cam E, Monnat J-Y. 2000 Apparent inferiority of first-time breeders in the kittiwake: the role of heterogeneity among age classes. *J. Anim. Ecol.* **69**, 380–394. (doi:10.1046/j.1365-2656.2000.00400.x)

8. Brommer JE, Pietiäinen H, Kolunen H. 1998 The effect of age at first breeding on Ural owl lifetime reproductive success and fitness under cyclic food conditions. *J. Anim. Ecol.* **67**, 359–369. (doi:10.1046/j.1365-2656.1998.00201.x)

9. Pradel R. 2005 Multievent: An Extension of Multistate Capture–Recapture Models to Uncertain States. *Biometrics* **61**, 442–447. (doi:10.1111/j.1541-0420.2005.00318.x)

10. Genovart M, Oro D, Tenan S. in press Immature survival, fertility and density-dependence drive global population dynamics in a long-lived species. *Ecology*

11. Choquet R, Rouan L, Pradel R. 2009 Program E-Surge: A Software Application for Fitting Multievent Models. In *Modeling Demographic Processes In Marked Populations* (eds DL Thomson, EG Cooch, MJ Conroy), pp. 845–865. Springer US.

12. Bradley JS, Gunn BM, Skira IJ, Meathrel CE, Wooller RD. 1999 Age-dependent prospecting and recruitment to a breeding colony of Short-tailed Shearwaters Puffinus tenuirostris. *Ibis* **141**, 277–285. (doi:10.1111/j.1474-919X.1999.tb07550.x)

13. Votier SC, Grecian WJ, Patrick S, Newton J. 2010 Inter-colony movements, at-sea behaviour and foraging in an immature seabird: results from GPS-PPT tracking, radio-tracking and stable isotope analysis. *Mar. Biol.* **158**, 355–362. (doi:10.1007/s00227-010-1563-9)

14. Henaux V, Bregnballe T, Lebreton J-D. 2007 Dispersal and recruitment during population growth in a colonial bird, the great cormorant Phalacrocorax carbo sinensis. *J. Avian Biol.* **38**, 44–57. (doi:10.1111/j.2006.0908-8857.03712.x)

15. Jenouvrier S, Tavecchia G, Thibault J-C, Choquet R, Bretagnolle V. 2008 Recruitment processes in long-lived species with delayed maturity: estimating key demographic parameters. *Oikos* **117**, 620–628. (doi:10.1111/j.0030-1299.2008.16394.x)

16. Fernández-Chacón A *et al.* 2013 When to stay, when to disperse and where to go: survival and dispersal patterns in a spatially structured seabird population. *Ecography* **36**, 1117–1126. (doi:10.1111/j.1600-0587.2013.00246.x)

17. Oro D, Pradel R. 1999 Recruitment of Audouin's gull to the Ebro Delta colony at metapopulation level in the western Mediterranean. *Mar. Ecol. Prog. Ser.* **180**, 267-273. (doi:10.3354/meps180267)

18. Schmidt KA, Dall SRX, Van Gils JA. 2010 The ecology of information: an overview on the ecological significance of making informed decisions. *Oikos* **119**, 304–316. (doi:10.1111/j.1600-0706.2009.17573.x)

19. Schmidt KA. 2004 Site fidelity in temporally correlated environments enhances population persistence. *Ecol. Lett.* **7**, 176–184. (doi:10.1111/j.1461-0248.2003.00565.x)

20. Danchin E, Boulinier T, Massot M. 1998 Conspecific reproductive success and breeding habitat selection: implications for the study of coloniality. *Ecology* **79**, 2415–2428. (doi:10.1890/0012-9658)

21. Pärt T, Doligez B. 2003 Gathering public information for habitat selection: prospecting birds cue on parental activity. *Proc. R. Soc. B* **270**, 1809–1813. (doi:10.1098/rspb.2003.2419)

22. Ward M. 2005 Habitat selection by dispersing yellow-headed blackbirds: evidence of prospecting and the use of public information. *Oecologia* **145**, 650–657. (doi:10.1007/s00442-005-0179-0)

23. Schjorring S, Gregersen J, Bregnballe T. 1999 Prospecting enhances breeding success of first-time breeders in the great cormorant, Phalacrocorax carbo sinensis. *Anim. Behav.* **57**, 647–654. (doi:10.1006/anbe.1998.0993)

24. Becker PH, Dittmann T, Ludwigs JD, Limmer B, Ludwig SC, Bauch C, Braasch A, Wendeln H. 2008 Timing of initial arrival at the breeding site predicts age at first reproduction in a long-lived migratory bird. *PNAS* **105**, 12349–12352. (doi:10.1073/pnas.0804179105)

25. Burkhalter JC, Fefferman NH, Lockwood JL. 2015 The impact of personality on the success of prospecting behavior in changing landscapes. *Curr. Zool.* **61**, 557–568. (doi:10.1093/czoolo/61.3.557)

26. Gil MA, Hein AM, Spiegel O, Baskett ML, Sih A. 2018 Social Information Links Individual Behavior to Population and Community Dynamics. *Trends Ecol. Evol.* **33**, 535–548. (doi:10.1016/j.tree.2018.04.010)

27. Kivelä SM, Seppänen J-T, Ovaskainen O, Doligez B, Gustafsson L, Mönkkönen M, Forsman JT. 2014 The past and the present in decision-making: the use of conspecific and heterospecific cues in nest site selection. *Ecology* **95**, 3428–3439. (doi:10.1890/13-2103.1)

28. Schmidt KA. 2017 Information thresholds, habitat loss and population persistence in breeding birds. *Oikos* **126**, 651–659. (doi:10.1111/oik.03703)

**Figure 1**. Conceptual model for recruitment of Audouin's gulls at the Ebro Delta with mean transition probabilities of interest between states from the selected model (see electronic supplementary material, S2) for each age class. Note that gulls only start breeding at 3y old. $NR_U$ represent those individuals that have never breed (i.e. not recruited) and not returned to the colony after fledging (i.e. uninformed). $NR_I$ represents those individuals informed that visit the colony at least once after fledging. R represents those individuals that recruit (i.e. start breeding at the study colony). Recruitment transitions are shown in blue. Confidence intervals (95%) for the estimates are shown in electronic supplementary material, S3.

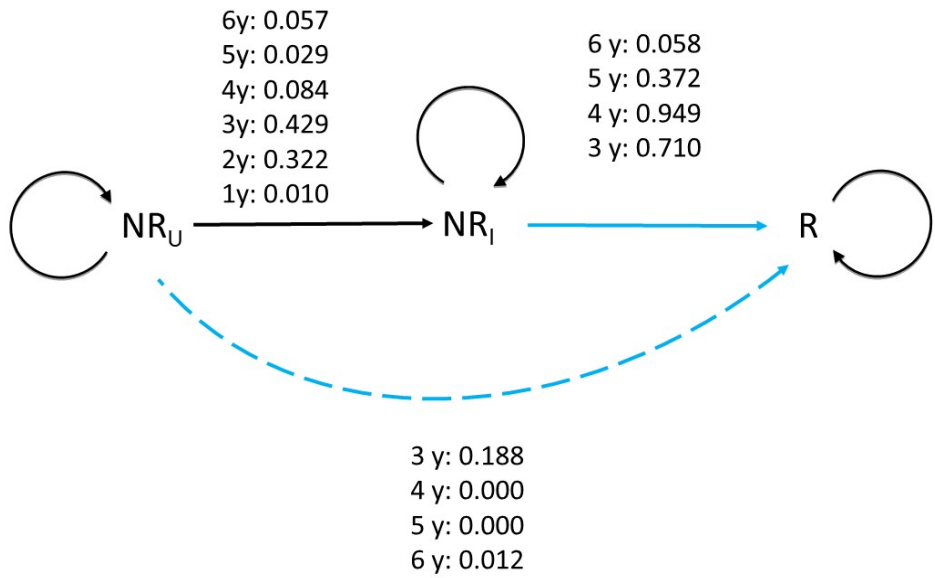